\documentstyle{article}

\author{M.\ De Francia
\and
H.\ Falomir\\
Departamento de F\'\i sica, Facultad de Ciencias Exactas,\\
Universidad Nacional de La Plata, \\
c.c. 67, 1900 La Plata, Argentina.
\and
M.\ Loewe\\
Facultad de F\'\i sica,\\
Pontificia Universidad Cat\'olica de Chile,\\
Casilla 306,
Santiago 22, Chile.
\and
E.\ M.\ Santangelo\\
Departamento de F\'\i sica, Facultad de Ciencias Exactas,\\
Universidad Nacional de La Plata,\\
c.c. 67, 1900
La Plata, Argentina.}
\title{Confined two-dimensional fermions at finite density
}

\def\NEG#1{{\rlap/#1}}
\def\tfrac#1#2{{\textstyle {#1 \over #2}}}
\def\dfrac#1#2{{\displaystyle {#1 \over #2}}}
\def\dprod{\mathop{\displaystyle \prod }}
\catcode`\@=11
\def\RIfM@{\relax\ifmmode}

\newif\iffirstchoice@
\firstchoice@true
\def\textfonti{\the\textfont\@ne}
\def\textfontii{\the\textfont\tw@}
\def\text{\RIfM@\expandafter\text@\else\expandafter\text@@\fi}
\def\text@@#1{\leavevmode\hbox{#1}}
\def\text@#1{\mathchoice
 {\hbox{\everymath{\displaystyle}\def\textfonti{\the\textfont\@ne}%
  \def\textfontii{\the\textfont\tw@}\textdef@@ T#1}}
 {\hbox{\firstchoice@false
  \everymath{\textstyle}\def\textfonti{\the\textfont\@ne}%
  \def\textfontii{\the\textfont\tw@}\textdef@@ T#1}}
 {\hbox{\firstchoice@false
  \everymath{\scriptstyle}\def\textfonti{\the\scriptfont\@ne}%
  \def\textfontii{\the\scriptfont\tw@}\textdef@@ S\rm#1}}
 {\hbox{\firstchoice@false
  \everymath{\scriptscriptstyle}\def\textfonti
  {\the\scriptscriptfont\@ne}%
  \def\textfontii{\the\scriptscriptfont\tw@}\textdef@@ s\rm#1}}}
\def\textdef@@#1{\textdef@#1\rm\textdef@#1\bf\textdef@#1\sl\textdef@#1\it}
\def\DN@{\def\next@}
\def\eat@#1{}
\def\textdef@#1#2{%
 \DN@{\csname\expandafter\eat@\string#2fam\endcsname}%
 \if S#1\edef#2{\the\scriptfont\next@\relax}%
 \else\if s#1\edef#2{\the\scriptscriptfont\next@\relax}%
 \else\edef#2{\the\textfont\next@\relax}\fi\fi}
\begin{document}

\maketitle
\begin{abstract}
We introduce the chemical potential in a system of two-dimensional massless
fermions, confined
to a finite region, by imposing twisted boundary conditions in the Euclidean
time direction. We explore in this simple model the application of functional
techniques which could be used in more complicated situations.

\end{abstract}

\newpage

\section{Introduction}

The thermodynamics of hadronic matter has recently received great attention
\cite{shuryak}, mainly in connection with the possible occurrence of a
deconfining phase transition at finite temperature. The difficulty of
studying this transition in the framework of QCD comes from the fact that
confinement has not been deduced in this theoretical context.

Consequently, effective models as, for example, the bag models \cite
{mit,chiral} have been introduced to mimic the confining properties of
strong
interactions. In these models, fields are confined to bounded regions and
subject to adequate boundary conditions. Finite volume effects turn out to
be relevant, and they have been studied in the thermodynamical limit, for
instance, in \cite{loewe}, and for a nonvanishing chemical
potential in \cite{mustafa}.

A complete analysis of the free energy for an MIT bag model at $T>0$, with $%
\mu =0$, has been performed in \cite{defrancia}. The corrections
introduced by chiral boundary conditions have also been studied \cite
{defrancia2}. Functional methods have been used in these papers to isolate
the finite temperature dependent pieces from the (divergent) Casimir energy.

The aim of the present paper is to study the possibility of introducing a
nonvanishing chemical potential through ``twisted" boundary conditions in
the
Euclidean time direction and treat it following the methods developed in
\cite{defrancia,pdet}. As a first approach to this problem we present here
the evaluation of the Gibbs free energy for the simple model of massless $1+1
$-dimensional fermions confined to a segment, for $T$ and $\mu \neq 0$.

Even though, for this toy model, the eigenvalue problem can be exactly solved,
we will rather follow an alternative approach. We will relate differences of
free energies of the system to the Green function satisfying adequate
(spatial and temporal)
boundary conditions. This approach is more likely useful in the realistic
four dimensional case, where eigenvalues cannot be explicitely solved for.
In section 4, we will reobtain the results in Section 2, making use of
the functional method developed in \cite{pdet}, which is based on the
evaluation of p-determinants.

\section{Two-dimensional fermions with $\mu \neq 0.$}

We consider a system of two-dimensional fermions confined in the segment $%
[0,L]$ and subject to given boundary conditions. As is well known, the
chemical potential can be introduced as an imaginary constant temporal gauge
field \cite{actor}. If the temperature is $T$
and the chemical potential is $\mu $, the Grand canonical partition function
can be expressed as
\begin{equation}
\label{pathint}
\begin{array}{c}
\Xi (T,L,\mu )=e^{-\beta G(T,L,\mu )} \\
\\
=\int
{\cal D}\bar \psi \ {\cal D}\psi \ e^{\int_0^1dt\int_0^1dx\bar \psi \left(
D(\beta ,L)-i\mu \gamma ^0\right) \psi } \\  \\
\sim Det\left( D(\beta ,L)-i\mu \gamma ^0\right) _{bc}
\end{array}
\end{equation}
where
\begin{equation}
D(\beta ,L)=\frac i\beta \gamma ^0\partial _t+\frac iL
\gamma ^1 \partial _x,\quad
\text{for\quad }0\leq t,x\leq 1,
\end{equation}
with
\begin{equation}
\gamma ^0=\left(
\begin{array}{cc}
0 & 1 \\
1 & 0
\end{array}
\right) ,\gamma ^1=\left(
\begin{array}{cc}
0 & -i \\
i & 0
\end{array}
\right) ,\gamma ^5=-i\gamma ^0\gamma ^1=\left(
\begin{array}{cc}
1 & 0 \\
0 & -1
\end{array}
\right)
\end{equation}
and $"bc"$ means that we must evaluate the functional determinant of the
differential operator defined on functions satisfying the boundary
conditions:
\begin{equation}
\label{BC}
\begin{array}{c}
B\psi (t,x)=0,
\text{ for }x=0,1, \\  \\
\psi (1,x)=-\psi (0,x).
\end{array}
\end{equation}
Here $B$ denotes an elliptic boundary condition to be satisfied at the
spatial edges.

So, we have the following relation for the Gibbs free energy:

\begin{equation}
\label{Dmu}
\begin{array}{c}
-\beta
\frac{\partial G}{\partial \mu }(\beta ,L,\mu )=Tr\{\frac \partial {\partial
\mu }\ln \left( D(\beta ,L)-i\mu \gamma ^0\right) _{bc}\} \\  \\
=Tr\{-i\gamma ^0K_{bc}(t,x;t^{\prime },x^{\prime })\},
\end{array}
\end{equation}
where $K_{bc}(t,x;t^{\prime },x^{\prime })$ is the Green function of the
problem,satisfying
\begin{equation}
\label{Green}
\begin{array}{c}
\left( D(\beta ,L)-i\mu \gamma ^0\right) K_{bc}(t,x;t^{\prime },x^{\prime
})=\delta (x-x^{\prime })\delta (t-t^{\prime }) \\
\\
BK_{bc}(t,x;t^{\prime },x^{\prime })=0,
\text{ for }x=0,1, \\  \\
K_{bc}(1,x;t^{\prime },x^{\prime })=-K_{bc}(0,x;t^{\prime },x^{\prime }).
\end{array}
\end{equation}
This can also be expressed in terms of
\begin{equation}
\begin{array}{c}
D(\beta ,L)-i\mu \gamma ^0=e^{\mu \beta t}D(\beta ,L)\ e^{-\mu \beta t}, \\
\\
K_{bc}(t,x;t^{\prime },x^{\prime })=e^{\mu \beta t}k(t,x;t^{\prime
},x^{\prime })\ e^{-\mu \beta t\prime },
\end{array}
\end{equation}
where $k(t,x;t^{\prime },x^{\prime })$ is the Green function of the problem
\begin{equation}
\begin{array}{c}
D(\beta ,L)\ k(t,x;t^{\prime },x^{\prime })=\delta (x-x^{\prime })\delta
(t-t^{\prime }) \\
\\
Be^{\mu \beta t}k(t,x;t^{\prime },x^{\prime })=0,
\text{ for }x=0,1, \\  \\
k(1,x;t^{\prime },x^{\prime })+e^{-\mu
\beta }k(0,x;t^{\prime },x^{\prime })=0.
\end{array}
\end{equation}
This function has the development
\begin{equation}
\label{desk}k(t,x;t^{\prime },x^{\prime })=L\sum\limits_{n=-\infty }^\infty
k_n(x,x^{\prime })\ e^{-i\Omega _n\beta (t-t^{\prime })},
\end{equation}
with the frequencies given by
\begin{equation}
\label{Omega}
\begin{array}{c}
\Omega _n=\omega _n-i\mu , \\
\\
\omega _n=\dfrac{(2n+1)\pi }\beta ,\text{ for }n\in {\bf Z.}
\end{array}
\end{equation}
The coefficients satisfy
\begin{equation}
k_n(x,x^{\prime })=e^{\gamma _5\Omega _nLx}k_n(0,x^{\prime })-i\gamma
^1e^{-\gamma _5\Omega _nL(x-x^{\prime })}H(x,x^{\prime }),
\end{equation}
with
\begin{equation}
H(x)=\left\{
\begin{array}{c}
1
\text{, for }x>0, \\ 0\text{, for }x<0.
\end{array}
\right.
\end{equation}
Now, we must impose the boundary conditions on the spatial edges. We will
adopt a static ``bag-like '' condition
\begin{equation}
\left( B\psi \right) (t,x)=(1+\NEG n)\psi (t,x)=0,\text{ for }x=0,1,
\end{equation}
such that, at $x=0$,
\begin{equation}
(1-\gamma ^1)k_n(0,x^{\prime })=\left(
\begin{array}{cc}
1 & i \\
-i & 1
\end{array}
\right) k_n(0,x^{\prime })=0.
\end{equation}
This implies that
\begin{equation}
k_n(0,x^{\prime })=\left(
\begin{array}{c}
1 \\
i
\end{array}
\right) \otimes \left(
\begin{array}{cc}
f(x^{\prime }) & g(x^{\prime })
\end{array}
\right) .
\end{equation}
On the other hand, at $x=1$,
\begin{equation}
(1+\gamma ^1)k_n(1,x^{\prime })=0,
\end{equation}
which means that
\begin{equation}
\left(
\begin{array}{cc}
1 & -i
\end{array}
\right) \left[ e^{\gamma _5\Omega _nLx}\left(
\begin{array}{c}
1 \\
i
\end{array}
\right) \otimes \left(
\begin{array}{cc}
f(x^{\prime }) & g(x^{\prime })
\end{array}
\right) -i\gamma ^1e^{-\gamma _5\Omega _nL(1-x\prime )}\right] =0.
\end{equation}
This gives
\begin{equation}
\label{greenfunc}
\begin{array}{c}
k(t,x;t^{\prime },x^{\prime })=iL\sum\limits_{n=-\infty }^\infty \left[
\dfrac{e^{\gamma _5\Omega _nLx}}{2\cosh (\Omega _nL)}\ (1+\gamma ^1)\
e^{-\gamma _5\Omega _nL(1-x\prime )}\right. \\  \\
\left. -\gamma ^1e^{-\gamma _5\Omega _nL(x-x\prime )}H(x-x^{\prime })\right]
\ e^{-i\Omega _n\beta (t-t^{\prime })}.
\end{array}
\end{equation}

So, we have for the derivative of the free energy
\begin{equation}
\label{dmug}
\begin{array}{c}
-\beta
\frac{\partial G}{\partial \mu }(\beta ,L,\mu )=Tr\{-i\gamma ^0e^{\mu \beta
t}k(t,x;t^{\prime },x^{\prime })\ e^{-\mu \beta t\prime }\} \\  \\
=L\int_0^1dx\int_0^1dt\ tr\left\{ \gamma ^0e^{\mu \beta (t-t^{\prime
})}\sum\limits_{n=-\infty }^\infty \left[
\dfrac{\ e^{-\gamma _5\Omega _nL(1-x-x\prime )}}{2\cosh (\Omega _nL)}\ \
\right. \right.  \\  \\
\left. \left. +\gamma ^1\ \left( \dfrac{e^{-\gamma _5\Omega _nL(1+x-x\prime
)}}{2\cosh (\Omega _nL)}-e^{-\gamma _5\Omega _nL(x-x\prime )}H(x-x^{\prime
})\right) \right] \ e^{-i\Omega _n\beta (t-t^{\prime })}\right\} \mid
_{(t^{\prime },x^{\prime })=(x,t)}.
\end{array}
\end{equation}
For $0<x,x^{\prime }<1,$ the first term in the series converges absolutely
and uniformly, even at $(t,x)=(t^{\prime },x^{\prime }).$ So, it can be
summed up in any order; in particular, one can first take the trace as a
matrix, obtaining a vanishing result. This is not the case for the second
term, containing the $\gamma ^1$-matrix. In fact,
\begin{equation}
\label{partsing}
\begin{array}{c}
\sum\limits_{n=-\infty }^\infty \left(
\dfrac{e^{-\gamma _5\Omega _nL(1+x-x\prime )}}{2\cosh (\Omega _nL)}%
-e^{-\gamma _5\Omega _nL(x-x\prime )}H(x-x^{\prime })\right) \ e^{-i\Omega
_n\beta (t-t^{\prime })}= \\  \\
\sum\limits_{n=0}^\infty \left[ \dfrac{e^{-\gamma _5\Omega _nL(1+x-x\prime )}
}{2\cosh (\Omega _nL)}-e^{-\gamma _5\Omega _nL(x-x\prime )}\left(
\begin{array}{cc}
0 & 0 \\
0 & 1
\end{array}
\right) \right] \ e^{-i\Omega _n\beta (t-t^{\prime })} \\
\\
+\sum\limits_{n=-\infty }^{-1}\left[ \dfrac{e^{-\gamma _5\Omega
_nL(1+x-x\prime )}}{2\cosh (\Omega _nL)}-e^{-\gamma _5\Omega _nL(x-x\prime
)}\left(
\begin{array}{cc}
1 & 0 \\
0 & 0
\end{array}
\right) \right] \ e^{-i\Omega _n\beta (t-t^{\prime })} \\
\\
+\sum\limits_{n=0}^\infty e^{-\gamma _5\Omega _nL(x-x\prime )}\left(
\begin{array}{cc}
-H(x-x^{\prime }) & 0 \\
0 & H(x^{\prime }-x)
\end{array}
\right) e^{-i\Omega _n\beta (t-t^{\prime })} \\
\\
+\sum\limits_{n=-\infty }^{-1}e^{-\gamma _5\Omega _nL(x-x\prime )}\left(
\begin{array}{cc}
H(x^{\prime }-x) & 0 \\
0 & -H(x-x^{\prime })
\end{array}
\right) \ e^{-i\Omega _n\beta (t-t^{\prime })},
\end{array}
\end{equation}
where the first two terms in the right hand side are absolutely convergent
even at $x^{\prime }=x$. For the last two terms in this equation we have,
for $x^{\prime }\neq x$,
\begin{equation}
\begin{array}{c}
S_{+}=\sum\limits_{n=0}^\infty e^{-\gamma _5\Omega _nL(x-x\prime )}\left(
\begin{array}{cc}
-H(x-x^{\prime }) & 0 \\
0 & H(x^{\prime }-x)
\end{array}
\right) e^{-i\Omega _n\beta (t-t^{\prime })}= \\
\\
\left(
\begin{array}{cc}
-H(x-x^{\prime })\dfrac{e^{\left( i\mu -\pi /\beta \right) \left[
L(x-x^{\prime })+i\beta \left( t-t^{\prime }\right) \right] }}{1-e^{-\tfrac{%
2\pi }\beta \left[ L(x-x^{\prime })L+i\beta \left( t-t^{\prime }\right)
\right] }} & 0 \\
0 & H(x^{\prime }-x)\dfrac{e^{\left( i\mu -\pi /\beta \right) \left[
L(x^{\prime }-x)L+i\beta \left( t-t^{\prime }\right) \right] }}{1-e^{-\tfrac{%
2\pi }\beta \left[ L(x^{\prime }-x)L+i\beta \left( t-t^{\prime }\right)
\right] }}
\end{array}
\right) ,
\end{array}
\end{equation}
and
\begin{equation}
\begin{array}{c}
S_{-}=\sum\limits_{n=-\infty }^{-1}e^{-\gamma _5\Omega _nL(x-x\prime
)}\left(
\begin{array}{cc}
H(x^{\prime }-x) & 0 \\
0 & -H(x-x^{\prime })
\end{array}
\right) \ e^{-i\Omega _n\beta (t-t^{\prime })}= \\
\\
\left(
\begin{array}{cc}
H(x^{\prime }-x)\dfrac{e^{-\left( i\mu +\pi /\beta \right) \left[ L(x^{\prime
}-x)-i\beta \left( t-t^{\prime }\right) \right] }}{1-e^{-\tfrac{2\pi }\beta
\left[ L(x^{\prime }-x)-i\beta \left( t-t^{\prime }\right) \right] }} & 0 \\
0 & -H(x-x^{\prime })\dfrac{e^{-\left( i\mu +\pi /\beta \right) \left[
L(x-x^{\prime })-i\beta \left( t-t^{\prime }\right) \right] }}{1-e^{-\tfrac{%
2\pi }\beta \left[ L(x-x^{\prime })-i\beta \left( t-t^{\prime }\right)
\right] }}
\end{array}
\right) ,
\end{array}
\end{equation}
which are singular at $(t^{\prime },x^{\prime })=(t,x).$ Calling $%
z=L(x-x^{\prime })+i\beta (t-t^{\prime }),\ \bar z=L(x-x^{\prime })-i\beta
(t-t^{\prime }),$ we have
\begin{equation}
\label{SMASS}
\begin{array}{c}
S_{+}+S_{-}= \\
\\
-\dfrac \beta {2\pi }\left(
\begin{array}{c}
\dfrac 1z+i\mu +\pi /\beta \left( H(x^{\prime }-x)-H(x-x^{\prime })\right)
+O(z/
{\beta }^2)\ ;\qquad 0 \\ 0\ ;\qquad \dfrac 1{\bar z}-i\mu +\pi /\beta \left(
H(x^{\prime }-x)-H(x-x^{\prime })\right) +O(\bar z/{\beta }^2)
\end{array}
\right) .
\end{array}
\end{equation}
Now, we can evaluate
\begin{equation}
\label{termlin}
\begin{array}{c}
L\ tr\left\{ \gamma ^0e^{\mu \beta (t-t^{\prime })}\gamma ^1\left(
S_{+}+S_{-}\right) \right\} = \\
\\
=-
\frac{iL\beta }{2\pi }e^{\mu \beta (t-t^{\prime })}\left\{ \dfrac 1z-\dfrac
1{\bar z}+2i\mu +O(\mid z\mid /{\beta }^2)\right\}  \\  \\
\begin{array}{c}
\\
\rightarrow  \\
_{t^{\prime }\rightarrow t}\
\end{array}
\dfrac{L\beta \mu }\pi +O(\mid x-x^{\prime }\mid L^2/{\beta })
\end{array}
\end{equation}
This correspond to the following (finite) contribution to $-\beta \frac{%
\partial G}{\partial \mu }(T,L,\mu ):$%
\begin{equation}
\label{contsing}
\begin{array}{c}
L\int_0^1dx\int_0^1dt\ tr\left\{ \gamma ^0e^{\mu \beta (t-t^{\prime
})}\gamma ^1\left( S_{+}+S_{-}\right) \right\} _{(t^{\prime },x^{\prime
})=(t,x)}= \\
\\
\dfrac \mu \pi L\beta .
\end{array}
\end{equation}
For the remaining terms in Eq.(\ref{partsing}), we can put $t^{\prime }=t$
and $x^{\prime }=x,$ and take the matrix trace inside the sum,

\begin{equation}
\begin{array}{c}
\ {L} \int_0^1 dx\int_0^1 dt\ tr\left\{ \gamma ^0\gamma
^1\sum\limits_{n=-\infty }^\infty \left[ \dfrac{e^{-\gamma _5\Omega _nL}}{%
2\cosh (\Omega _nL)}-\left(
\begin{array}{cc}
H(-n-1/2) & 0 \\
0 & H(n+1/2)
\end{array}
\right) \right] \ \right\} \\
\\
=-iL\sum\limits_{n=-\infty }^\infty \left\{ \tanh (\Omega
_nL)-sign(n+1/2)\right\} .
\end{array}
\end{equation}
We finally get
\begin{equation}
-\beta \frac{\partial G}{\partial \mu }(\beta ,L,\mu )=\dfrac \mu \pi
L\beta -iL\ \sum\limits_{n=-\infty }^\infty \left\{ \tanh \left[ \left(
\omega _n-i\mu \right) L\right] -sign(n+1/2)\right\} .
\label{numero}
\end{equation}
Notice that, while the first term in the r.h.s. of (\ref{numero}),
coming from the singular behaviour of the Green function,
is linear in $\mu $, the second one, a $\pi$-periodic function
of $\mu L$, contains the finite size effects (vanishing for $L
\rightarrow \infty$). It is easy to see from this expression that,
in the $\beta \rightarrow \infty$ limit, the mean value of
the particle number is
\begin{equation}
\lim _{\beta \rightarrow \infty} \bar{N}=
\left[ {{\mu L}\over{\pi }} \right] ,
\end{equation}
where $[x]$ means the integer part of $x$.

Now, (\ref{numero}) can be integrated to obtain:
\begin{equation}
\label{egibbs}
\begin{array}{c}
-\beta \left\{ \ G(\beta ,L,\mu )-G(\beta ,L,0)\right\} = \\
\\
\dfrac{\mu ^2}{2\pi }L\beta +\ \sum\limits_{n=-\infty }^\infty \ln \
\left\{ \dfrac{\cosh \left[ \left( \omega _n-i\mu \right) L\right] }{\cosh
\left[ \omega _nL\right] }+i\mu \ L\ sign(n+1/2)\right\} .
\end{array}
\end{equation}
Taking into account that $\omega _{-n}=-\omega _{(n-1)}$ we have, for the
piece of the free energy which depends on $\mu $, the (finite) result
\begin{equation}
\label{elgibbs}
\begin{array}{c}
-\beta \left\{ \ G(\beta ,L,\mu )-G(\beta ,L,0)\right\} = \\
\\
=
\dfrac{\mu ^2}{2\pi }L\beta +\ \ln \ \left\{ \dprod\limits_0^\infty \dfrac{%
\left( 1+2\cos (2\mu L)e^{-2\omega _nL}+e^{-4\omega _nL}\right) }{\left(
1+2e^{-2\omega _nL}+e^{-4\omega _nL}\right) }\right\} \\  \\
=\dfrac{\mu ^2}{2\pi }L\beta +\ln \left\{ \dfrac{\theta _3(\mu L,e^{-2\pi
L/\beta })}{\theta _3(0,e^{-2\pi L/\beta })}\right\} ,
\end{array}
\end{equation}
where $\theta _3 (u,q)$ is the Jacobi theta function \cite{tabla}.

%%%%%%%%%%%%%%%%%%%%%%%%%%%%%%%%%%%%%%%%%%%%%%%%%%%%%%%%%%%%%%%%%%%%%%%

\section{The free energy for $\mu =0$}

In order to have the complete Gibbs function, we will evaluate in this
section the free energy of our system of two-dimensional fermions as a
function of temperature, for $\mu =0.$ We take
\begin{equation}
\label{dbeta}\frac \partial {\partial \beta }\ln \ Det\left( D(\beta
,L)\right) _{bc}=Tr\left\{ \frac{-i}{\beta ^2}\gamma ^0\partial
_tk(t,x;t^{\prime },x^{\prime })\right\} ,
\end{equation}
where the Green function of the operator is given by (\ref{greenfunc}) with $%
\mu =0$
\begin{equation}
\begin{array}{c}
k(t,x;t^{\prime },x^{\prime })=iL\sum\limits_{n=-\infty }^\infty \left[
\dfrac{e^{\gamma _5\omega _nLx}}{2\cosh (\omega _nL)}\ (1+\gamma ^1)\
e^{-\gamma _5\omega _nL(1-x\prime )}\right. \\  \\
\left. -\gamma ^1e^{-\gamma _5\omega _nL(x-x\prime )}H(x-x^{\prime })\right]
\ e^{-i\omega _n\beta (t-t^{\prime })}.
\end{array}
\end{equation}
As in Section 1, the term in the r.h.s. of Eq.(\ref{dbeta}) which does not
contain $\gamma ^1$ gives a vanishing contribution. For the remaining, we
get two contributions to the trace: The first, coming from the absolutely
convergent series (for $0<x,x^{\prime }<1)$%
\begin{equation}
\begin{array}{c}
\dfrac L\beta Tr\left\{ \gamma _5\sum\limits_{n=-\infty }^\infty \omega
_ne^{-i\omega _n\beta (t-t^{\prime })}e^{-\gamma _5\omega _nL(x-x\prime
)}\left[ \dfrac{e^{-\gamma _5\omega _nL}}{2\cosh (\omega _nL)}-\left(
\begin{array}{c}
H(-n-1/2);\quad 0 \\
0;\quad H(n+1/2)
\end{array}
\right) \right] \right\} \\
\\
=-\dfrac L\beta \sum\limits_{n=-\infty }^\infty \omega _n\left\{ \tanh
(\omega _nL)-sign(n+1/2)\right\} \\
\\
=2\dfrac \partial {\partial \beta }\sum\limits_{n=0}^\infty \ln \left\{ 1+\
e^{-2\omega _nL}\right\} .
\end{array}
\end{equation}
The second one turns out to be divergent, and therefore a
regularization procedure is required. We will use
a ``point-splitting" prescription to get
\begin{equation}
\begin{array}{c}
\dfrac{iL}{\beta ^2}Tr\left\{ \gamma _5\ \partial _t\sum\limits_{n=-\infty
}^\infty e^{-i\omega _n\beta (t-t^{\prime })}e^{-\gamma _5\omega
_nL(x-x\prime )}\right. \times \\  \\
\left. \left(
\begin{array}{c}
H(x^{\prime }-x)H(-n-1/2)-H(x-x^{\prime })H(n+1/2);\qquad 0 \\
\\
0;\qquad H(x^{\prime }-x)H(n+1/2)-H(x-x^{\prime })H(-n-1/2)
\end{array}
\right) \right\} \\
\\
=2\dfrac L{\beta ^2}\int_0^1dx\int_0^1dt\
{\it Im\ }\partial _t\left\{ H(x-x^{\prime })\ \sum\limits_{n=0}^\infty
e^{-(2n+1)\frac \pi \beta \left[ L(x-x^{\prime })+i\beta (t-t^{\prime
})\right] }\right. \\  \\
\left. +H(x^{\prime }-x)\ \sum\limits_{n=0}^\infty e^{-(2n+1)\frac \pi \beta
\left[ L(x^{\prime }-x)+i\beta (t-t^{\prime })\right] }\right\} \mid
_{(t^{\prime },x^{\prime })=(t,x)} \\
\\
=\dfrac L{\beta ^2}\int_0^1dx\int_0^1dt\
{\it Im\ }\partial _t\left\{ \dfrac 1{\sinh \left( \dfrac \pi \beta \left[
L\mid x-x^{\prime }\mid +i\beta (t-t^{\prime })\right] \right) }\right\} \
\mid _{(t^{\prime },x^{\prime })=(t,x)} \\  \\
\begin{array}{c}
\\
\rightarrow \\
t^{\prime }\rightarrow t
\end{array}
-\dfrac{\pi L}{\beta ^2}\dfrac{\cosh (\pi \epsilon /\beta )}{\left[ \sinh
(\pi \epsilon /\beta )\right] ^2}\ =\dfrac \partial {\partial \beta }\left\{
-\dfrac{L\beta }{\pi \epsilon ^2}-\dfrac{\pi L}{3\beta }+O(\epsilon
^2)\right\} ,
\end{array}
\end{equation}
for $\epsilon =L\mid x^{\prime }-x\mid <<\beta ,$ which shows a singular
piece proportional to $L.$

So, we have for the free energy at $\mu =0$%
\begin{equation}
\begin{array}{c}
\ F(\beta ,L)=-\frac 2\beta \sum\limits_{n=0}^\infty \ln \left\{ 1+\
e^{-2\omega _nL}\right\} +\dfrac L{\pi \epsilon ^2}-
\dfrac{\pi L}{3\beta ^2}+\dfrac C\beta \\  \\
=-\dfrac 1\beta \ln \left\{ \theta _3(0,e^{-\dfrac{2\pi L}\beta })\right\}
+\dfrac 1\beta \sum\limits_{n=1}^\infty \ln \left( 1-e^{-\dfrac{4\pi Ln}%
\beta }\right) +\dfrac L{\pi \epsilon ^2}-\dfrac{\pi L}{3\beta ^2}+\dfrac
C\beta ,
\end{array}
\end{equation}
with $C$ an integration constant.

The Casimir energy is obtained through the limit of this expression for
vanishing temperature,
\begin{equation}
\begin{array}{c}
\lim _{\beta \rightarrow \infty }F(\beta ,L)=E_{Cas.}(L)=-\frac 1\pi
\int_0^\infty d\omega \ln \left\{ 1+\ e^{-2\omega L}\right\} +\dfrac L{\pi
\epsilon ^2} \\
\\
=-\dfrac \pi {24L}+\dfrac L{\pi \epsilon ^2},
\end{array}
\end{equation}
where one can see that the singular $O(\epsilon ^{-2})$ part can be
eliminated through a renormalization of the vacuum energy density. The
finite part of $E_{Cas.}(L),$ $-\dfrac \pi {24L},$ coincides with the
result
obtained in ref.(\cite{ECAS}), and gives rise to an attractive force
between
the edges of the segment where the fermions are confined.

Then, we get the finite result
\begin{equation}
\label{F-Ecas}
\begin{array}{c}
F(\beta ,L)-E_{Cas.}(L)= \\
\\
-\dfrac 1\beta \ln \left\{ \theta _3(0,e^{-\dfrac{2\pi L}\beta })\right\}
+\dfrac 1\beta \sum\limits_{n=1}^\infty \ln \left( 1-e^{-\dfrac{4\pi Ln}%
\beta }\right) -\dfrac{\pi L}{3\beta ^2}+\dfrac C\beta +\dfrac \pi {24L}.
\end{array}
\end{equation}
The undetermined constant $C$ can be evaluated by imposing the vanishing of
entropy at zero temperature. Since
\begin{equation}
\label{entropy}S=\beta ^2\frac \partial {\partial \beta }F(\beta ,L),
\end{equation}
for $\beta \rightarrow \infty $ the Poisson formula for the previous series
allows us to write
\begin{equation}
S=\beta ^2\frac \partial {\partial \beta }\left\{ \dfrac C\beta +O(\beta
^{-2})\right\} =-C+O(\beta ^{-1}),
\end{equation}
implying that $C=0.$

We finally get for the free energy as a function of $\beta ,L,\mu :$%
\begin{equation}
\label{rfinal}
\begin{array}{c}
\ G(\beta ,L,\mu )-\ E_{Cas.}(L)= \\
\\
=-\dfrac{\mu ^2L}{2\pi }-\dfrac 1\beta \ln \left\{ \theta _3(\mu L,e^{-
\tfrac{2\pi L}\beta })\right\} +\dfrac 1\beta \sum\limits_{n=1}^\infty \ln
\left( 1-e^{-\tfrac{4\pi Ln}\beta }\right) -\dfrac{\pi L}{3\beta ^2}+\dfrac
\pi {24L}.
\end{array}
\end{equation}

%%%%%%%%%%%%%%%%%%%%%%%%%%%%%%%%%%%%%%%%%%%%%%%%%%%%%%%%%%%%%%%%%%%%%%%%%%

\section{The chemical potential through a boundary value problem}

In this section, we will show how the result in Section 1 can alternatively
be obtained with the methods developed in ref.\cite{pdet}, by introducing
the
chemical potential through suitable boundary conditions.

As before, we will consider the differential operator
\begin{equation}
\label{oper}D(\beta ,L)=\frac i\beta \gamma ^0\partial _t+\frac iL\gamma
^1\partial _x,
\end{equation}
with the dimensionless variables $t,x$ taking values in the segment $[0,1].$
We will take $D(\beta ,L)$ to act on differentiable functions satisfying the
boundary conditions
\begin{equation}
\label{AyB}
\begin{array}{c}
\left( A(\eta )\chi \right) (x)\equiv \chi (t=1,x)+e^{i\eta }\chi (t=0,x)=0,
\\
\\
\left( B\chi \right) (t,x)\equiv [1+\NEG n]\chi (t,x)=0,\text{ for }x=0,1,
\end{array}
\end{equation}
and call this operator $\left( D(\beta ,L)\right) _{A(\eta ),B}.$

The method requires a basis in the kernel of the differential operator,
\linebreak
$\ker \ D(\beta ,L).$ The calculation can be greatly simplified by choosing
a complete system of functions satisfying the boundary conditions at the
spatial edges. Such basis can be constructed from the eigenfunctions
of the Hermitian operator $\left( -i\gamma _5\frac d{dx}\right) _B\ $in $%
[0,1]$:
\begin{equation}
\begin{array}{c}
-i\gamma _5\frac d{dx}\chi _n(x)=\lambda _n\chi _n(x), \\
\\
\left( B\chi _n\right) (0)=0=\left( B\chi _n\right) (1),
\end{array}
\end{equation}
which are given by
\begin{equation}
\label{base}\chi _n(x)=e^{i\pi (n+1/2)x\gamma _5}\left(
\begin{array}{c}
1 \\
i
\end{array}
\right) \text{, with }\lambda _n=\pi (n+1/2),\ n\in {\cal Z.}
\end{equation}
Now, a basis in $\ker \ D(\beta ,L)$ is
\begin{equation}
\label{baseker}\psi _n(t,x)=e^{(n+1/2)\frac \pi L\left[ \beta t+iLx\gamma
_5\right] }\left(
\begin{array}{c}
1 \\
i
\end{array}
\right) \text{, with }\ n\in {\cal Z.}
\end{equation}
The following step is to get the projected boundary values of $\psi _n(t,x)$
through the boundary operators $A(\eta )$ and $B$:
\begin{equation}
\label{proyeccion}
\begin{array}{c}
H_n(t,x;\eta )=\left(
\begin{array}{c}
\left( A(\eta )\psi _n\right) (x) \\
\left( B\psi _n\right) (t,0) \\
\left( B\psi _n\right) (t,1)
\end{array}
\right) =\left(
\begin{array}{c}
h_n(x;\eta ) \\
0 \\
0
\end{array}
\right) , \\
\\
h_n(x;\eta )=\psi _n(t=1,x)+e^{i\eta }\psi _n(t=0,x)= \\
\\
=\left[ e^{(n+1/2)\pi \beta /L}+e^{i\eta }\right] e^{i\pi (n+1/2)x\gamma
_5}\left(
\begin{array}{c}
1 \\
i
\end{array}
\right) .
\end{array}
\end{equation}
Forman's operator, $\tilde \Phi (\eta ^{\prime },\eta )$, is defined as
\begin{equation}
\label{forman}H_n(t,x;\eta ^{\prime })=\tilde \Phi (\eta ^{\prime },\eta )\
H_n(t,x;\eta ),
\end{equation}
for any basis in $\ker \ D(\beta ,L).$ Since the operator $B$ does not
depends on the parameter $\eta ,$ $\tilde \Phi (\eta ^{\prime },\eta )$ has
the form
\begin{equation}
\tilde \Phi (\eta ^{\prime },\eta )=\left(
\begin{array}{cc}
\Phi (\eta ^{\prime },\eta ) & \Phi ^{\prime }(\eta ^{\prime },\eta ) \\
0 & Id_{2\times 2}
\end{array}
\right) ,
\end{equation}
and our election of basis allows us to determine $\Phi (\eta ^{\prime },\eta
)$ from
\begin{equation}
\begin{array}{c}
h_n(x;\eta ^{\prime })=\Phi (\eta ^{\prime },\eta )\ h_n(x;\eta )= \\
\\
=\left( \dfrac{e^{(n+1/2)\pi \beta /L}+e^{i\eta \prime }}{e^{(n+1/2)\pi
\beta /L}+e^{i\eta }}\right) h_n(x;\eta ).
\end{array}
\end{equation}
Note that the $h_n(x;\eta )\sim \chi _n(x)$, for $n\in {\cal Z,}$ constitute
a complete system in {\cal L$^2(0,1),$} and $\Phi (\eta ^{\prime },\eta )$
is diagonal in this basis:
\begin{equation}
\label{FI}\left( \Phi (\eta ^{\prime },\eta )\right) _{n,m}=\delta
_{n,m}\left( \dfrac{1+e^{i\eta \prime -(n+1/2)\pi \beta /L}}{1+e^{i\eta
-(n+1/2)\pi \beta /L}}\right) .
\end{equation}
In what follows, we will need the quotient of $\Phi $'s for two sets of
parameters, $(\beta ,L)$ and $(\beta _0,L_0)$ respectively,
\begin{equation}
\label{cofi}\left( \Phi (\eta ^{\prime },\eta )\ \Phi _0^{-1}(\eta ^{\prime
},\eta )\right) _{n,m}=\delta _{n,m}\left( \dfrac{1+e^{i\eta \prime
-(n+1/2)\pi \beta /L}}{1+e^{i\eta -(n+1/2)\pi \beta /L}}\right) \left( \frac{%
1+e^{i\eta -(n+1/2)\pi \beta _0/L_0}}{1+e^{i\eta \prime -(n+1/2)\pi \beta
_0/L_0}}\right) .
\end{equation}
It is easy to see that $\left[ \Phi (\eta ^{\prime },\eta )\ \Phi
_0^{-1}(\eta ^{\prime },\eta )-1\right] $ is a trace class operator, so the
determinant $\det _1\left( \Phi (\eta ^{\prime },\eta )\ \Phi _0^{-1}(\eta
^{\prime },\eta )\right) $ exists:
\begin{equation}
\label{det1}
\begin{array}{c}
\det \nolimits_1\left( \Phi (\eta ^{\prime },\eta )\ \Phi _0^{-1}(\eta
^{\prime },\eta )\right) =\dprod\limits_0^\infty \left|
\dfrac{1+e^{i\eta \prime -(n+1/2)\pi \beta /L}}{1+e^{i\eta -(n+1/2)\pi \beta
/L}}\right| ^2\left| \dfrac{1+e^{i\eta -(n+1/2)\pi \beta _0/L_0}}{1+e^{i\eta
\prime -(n+1/2)\pi \beta _0/L_0}}\right| ^2 \\  \\
=\dfrac{\theta _3(\frac{\eta ^{\prime }}2,e^{-\tfrac{\pi \beta }{2L}})}{%
\theta _3(\frac \eta 2,e^{-\tfrac{\pi \beta }{2L}})}\ \dfrac{\theta _3(\frac
\eta 2,e^{-\tfrac{\pi \beta _0}{2L_0}})}{\theta _3(\frac{\eta ^{\prime }}%
2,e^{-\tfrac{\pi \beta _0}{2L_0}})}.
\end{array}
\end{equation}
The result established in ref.\cite{pdet} is
\begin{equation}
\begin{array}{c}
\det \nolimits_1\left\{ \left( D(\beta ,L)\right) _{A(\eta ^{\prime }),B}\
\left( D(\beta _0,L_0)\right) _{A(\eta ^{\prime }),B}^{-1}\ \left( D(\beta
_0,L_0)\right) _{A(\eta ),B}\ \left( D(\beta ,L)\right) _{A(\eta
),B}^{-1}\right\} = \\
\\
\det \nolimits_1\left( \Phi (\eta ^{\prime },\eta )\ \Phi _0^{-1}(\eta
^{\prime },\eta )\right) .
\end{array}
\end{equation}
Now, we choose the parameter $\eta ^{\prime }=i\mu ^{\prime }\beta ,\ \eta
=i\mu \ \beta $ and noticing that
\begin{equation}
\begin{array}{c}
\left( D(\beta ,L)\right) _{A(\eta ),B}=\left( e^{-i\eta t}D(\beta
,L)e^{i\eta t}\right) _{A(0),B} \\
\\
=\left( \frac i\beta \gamma ^0\partial _t+\frac iL\gamma ^1\partial _x-
\frac
\eta \beta \gamma ^0\right) _{A(0),B}\ ,
\end{array}
\end{equation}
we can write
\begin{equation}
\begin{array}{c}
-\beta \ G(\beta ,L,\mu ^{\prime })+\beta _0G(\beta _0,L_0,\mu ^{\prime
}\beta /\beta _0)-\beta _0\ G(\beta _0,L_0,\mu \ \beta /\beta _0)+\beta \
G(\beta ,L,\mu ) \\
\\
=\ln \left\{ \dfrac{\theta _3(\frac{i\mu ^{\prime }\beta }2,e^{-\tfrac{\pi
\beta }{2L}})}{\theta _3(\frac{i\mu \ \beta }2,e^{-\tfrac{\pi \beta }{2L}})}\
\dfrac{\theta _3(\frac{i\mu \ \beta }2,e^{-\tfrac{\pi \beta _0}{2L_0}})}{%
\theta _3(\frac{i\mu ^{\prime }\beta }2,e^{-\tfrac{\pi \beta _0}{2L_0}})}%
\right\} .
\end{array}
\end{equation}
For $\mu ^{\prime }=0$ this reduces to
\begin{equation}
\begin{array}{c}
\beta \ \left[ G(\beta ,L,\mu )-\ G(\beta ,L,0)\right] -\beta _0\ \left[
G(\beta _0,L_0,\mu \ \beta /\beta _0)-G(\beta _0,L_0,0)\right] \\
\\
=\ln \left\{ \dfrac{\theta _3(0,e^{-\tfrac{\pi \beta }{2L}})}{\theta _3(
\frac{i\mu \ \beta }2,e^{-\tfrac{\pi \beta }{2L}})}\
\dfrac{\theta _3(\frac{%
i\mu \ \beta }2,e^{-\tfrac{\pi \beta _0}{2L_0}})}{\theta _3(0,e^{-\tfrac{\pi
\beta _0}{2L_0}})}\right\} ,
\end{array}
\end{equation}
and we are still free to choose the parameters $\beta _0$ and $L_0$
conveniently.

Notice that, for $\beta /\beta _0<<1,$ we can write
\begin{equation}
\begin{array}{c}
\beta _0\ \left[ G(\beta _0,L_0,\mu \ \beta /\beta _0)-G(\beta
_0,L_0,0)\right] =\beta _0
\dfrac{\mu \ \beta }{\beta _0}\dfrac{\partial G}{\partial \mu _0}(\beta
_0,L_0,\mu _0) \\ =-\mu \ \beta \ \bar N(\beta _0,L_0,\mu _0),
\end{array}
\end{equation}
for $0<$ $\mu _0<\mu \ \beta /\beta _0$, and $\bar N(\beta _0,L_0,\mu _0)$
the mean value of the particle number. So
\begin{equation}
\beta _0\ \left[ G(\beta _0,L_0,\mu \ \beta /\beta _0)-G(\beta
_0,L_0,0)\right]
\begin{array}{c}
\\
\rightarrow  \\
\beta _0\rightarrow \infty
\end{array}
0.
\end{equation}
On the other hand
\begin{equation}
\lim _{\beta _0\rightarrow \infty }\ {{\theta _3(\frac{i\mu \ \beta }2,
e^{-
\tfrac{\pi \beta _0}{2L_0}})}\over{\theta _3(0,e^{-
\tfrac{\pi \beta _0}{2L_0}})}}=1,
\end{equation}
for any value of $\frac{\mu \ \beta }2$. Then, we have
\begin{equation}
\begin{array}{c}
\beta \ \left[ G(\beta ,L,\mu )-\ G(\beta ,L,0)\right] =\ln \left\{
\dfrac{\theta _3(0,e^{-\tfrac{\pi \beta }{2L}})}{\theta _3(\frac{i\mu \
\beta
}2,e^{-\tfrac{\pi \beta }{2L}})}\ \right\}  \\  \\
=\dfrac{-\mu ^2\beta L}{2\pi }+\ln \left\{ \dfrac{\theta _3(0,
e^{-\tfrac{2\pi
L}\beta })}{\theta _3(\mu L,e^{-\tfrac{2\pi L}\beta })}\ \right\} ,
\end{array}
\end{equation}
\newpage
where we have used the inversion formula for the Jacobi $\theta _3$%
-function,
\begin{equation}
\theta _3(\frac{i\mu \ \beta }2,e^{-\tfrac{\pi \beta }{2L}})=e^{\tfrac{\mu
^2\beta L}{2\pi }}\theta _3(\mu L,e^{-\tfrac{2\pi L}\beta }).
\end{equation}
This result coincides with the one obtained in Eq.(\ref{elgibbs}).

\section{Conclusions}

In this paper, we studied the possibility of simulating the presence
of a chemical potential by imposing ``twisted" boundary conditions in the
Euclidean time direction.
This mechanism allowed us to relate the free energy of a system of
massless confined
fermions to a functional determinant of the Dirac operator.

To study such determinant, we made use of two approaches. The first one
is based on the knowledge of the Green function satisfying adequate
boundary conditions. The other relies on functional techniques \cite{pdet}
which allow to perform the calculation starting from boundary values of
functions in the kernel of the Dirac operator.

In Section 2, the difference between Gibbs free energies with and
without chemical
potential was written in terms of the $\mu$-integral of a trace
involving the Green function.
A careful treatment of this trace (which is not absolutely
convergent), through point splitting regularization, put in evidence
the presence of a $\mu ^2$-dependent term.

For completeness, in section 3, the Helmholtz free energy was
computed by a similar method, employing this time the $\beta $-derivative.
The constant of integration was fixed from the vanishing of entropy
at zero temperature. Moreover, the $\beta \rightarrow \infty$
limit allowed us to identify the Casimir energy of the system.

Although the results of sections 2 and 3, i.e. the Gibbs free energy of this
simple
model, could also have been obtained directly from the eigenvalues of the
Dirac Hamiltonian
(which, in this case, can be exactly evaluated), the method developed in
those sections can be applied to more complicated and realistic situations,
e.g. the $3+1$-dimensional bag model.

In section 4, we made an alternative calculation following the techniques
introduced in \cite{pdet}. Starting from an adequately selected basis in the
kernel of the
Dirac operator, the Fredholm determinant of quotients of Forman's operator
could be easily constructed.
Through an analytic extension in the parameter defining the twisted temporal
boundary condition, we reobtained the $\mu $-dependent piece of
the Gibbs free energy of the system.

The calculational scheme explored in the present paper should allow for the
introduction of a nonvanishing chemical potential in the $3+1$-dimensional
chiral bag model, thus complementing the results in
\cite{defrancia,defrancia2}.

\newpage

{\bf Acknowledgements}

This work was supported in part by Fundaciones Andes y Antorchas
under Contract 12345/9, CO\-NICET (Argentina) and
FONDECyT (Chile) under Grant 0751/92.

\end{document}